RESEARCH ARTICLE

# EyeAI: AI-Assisted Ocular Disease Detection for Equitable Healthcare Access

Shiv Garg (student)[1 *], Dr. Ginny Berkemeier (mentor)[1]

1. FCS Innovation Academy, 125 Milton Ave., Alpharetta, Georgia, 30009, USA

## Abstract

Ocular disease affects billions of individuals unevenly worldwide. It continues to increase in prevalence with trends of growing populations of diabetic people, increasing life expectancies, decreasing ophthalmologist availability, and rising costs of care. We present EyeAI, a system designed to provide artificial intelligence-assisted detection of ocular diseases, thereby enhancing global health. EyeAI utilizes a convolutional neural network model trained on 1,920 retinal fundus images to automatically diagnose the presence of ocular disease based on a retinal fundus image input through a publicly accessible web-based application. EyeAI performs a binary classification to determine the presence of any of 45 distinct ocular diseases, including diabetic retinopathy, media haze, and optic disc cupping, with an accuracy of 80%, an AUROC of 0.698, and an F1-score of 0.8876. EyeAI addresses barriers to traditional ophthalmologic care by facilitating low-cost, remote, and real-time diagnoses, particularly for equitable access to care in underserved areas and for supporting physicians through a secondary diagnostic opinion. Results demonstrate the potential of EyeAI as a scalable, efficient, and accessible diagnostic tool. Future work will focus on expanding the training dataset to enhance the model's accuracy further and improve its diagnostic capabilities.

## Keywords

## Introduction

An estimated 2.2 billion people are visually impaired worldwide, with over 1 billion cases that were preventable if diagnosed and treated earlier.[1] Furthermore, the proportion of people affected by ocular disease continues to increase. This increase is attributed to four primary factors: diabetes, age, the ophthalmologist's availability, and cost.

### *Diabetes*

The number of people with diabetes is increasing, with a projected 643 million individuals in 2030 being diabetic.[2] Diabetes is a significant factor that contributes to ocular disease; diabetic retinopathy, a major cause of blindness in adults, is often a complication of diabetes. Additionally, more than 75% of diabetic adults live in low to middle-income countries, highlighting the importance of low-cost ocular disease diagnosis.[3] An expected 48% increase in the number of individuals affected by diabetic retinopathy by 2030 demonstrates the rapid proliferation of ocular disease as a result of the growing diabetic population.[4] A novel system of rapid disease detection is vital.

### *Age*

Additionally, the detection of ocular diseases is especially essential for the elderly population. After the age of 40, the risk of ocular disease increases significantly, and by the age of 75, visual impairment becomes extremely prevalent.[5] The proportion of people aged over 60 years is expected to nearly double globally between 2015 and 2050, highlighting the trend of an increasingly elderly population.[6] This elderly population is at a heightened risk of ocular disease, further underscoring the need for an efficient diagnostic solution. Additionally, in 2050, 80% of elderly people will be living in low and middle-income communities, underscoring the importance of low-cost ocular disease diagnoses.[6]

### *Ophthalmologist Availability*

Many people cannot receive a diagnosis of ocular disease before complications arise. There is a general lack of access to ophthalmologist care for diagnosing ocular diseases, as the estimated mean global ophthalmologist density is 31.7 per million people.[7] In addition, ophthalmologist density is not evenly distributed between urban and rural areas, as well as between low-income and high-income communities. On average, there are 3.7 ophthalmologists per million people in low-income countries, while there are 76.2 ophthalmologists per million people in high-income countries.[7] As ophthalmologists move to urban, high-income areas to earn higher salaries, there is a growing population with reduced access to ophthalmologic care. A low-cost and efficient solution for ocular disease diagnosis is necessary to address the discrepancy in ophthalmologist availability.

### *Cost*

People in lower-income areas often have limited access to ophthalmologists, as many facilities lack the necessary resources to provide accurate diagnoses. Appointments with an ophthalmologist can cost up to 200 USD, which further impedes the accessibility of ophthalmologic care for many individuals. Currently, diagnosis of ocular disease requires an ophthalmologist, which costs 320,000 USD on average per year. In many areas, this is not an affordable option. However, retinal imaging machines used to image retinal fundi do not require specialized training to operate, and functional machines can cost approximately 5,000 USD on average, making them a significantly more affordable investment for a community. The lack of affordability of traditional ophthalmologic care highlights the need for a cost-effective, efficient solution for ocular disease diagnosis.

### *EyeAI*

Machine learning has immense potential to identify pathological features in retinal fundi, enabling informed diagnoses of ocular diseases. To provide a low-touch, low-cost medical service for all, the EyeAI System was developed (Figure 1).

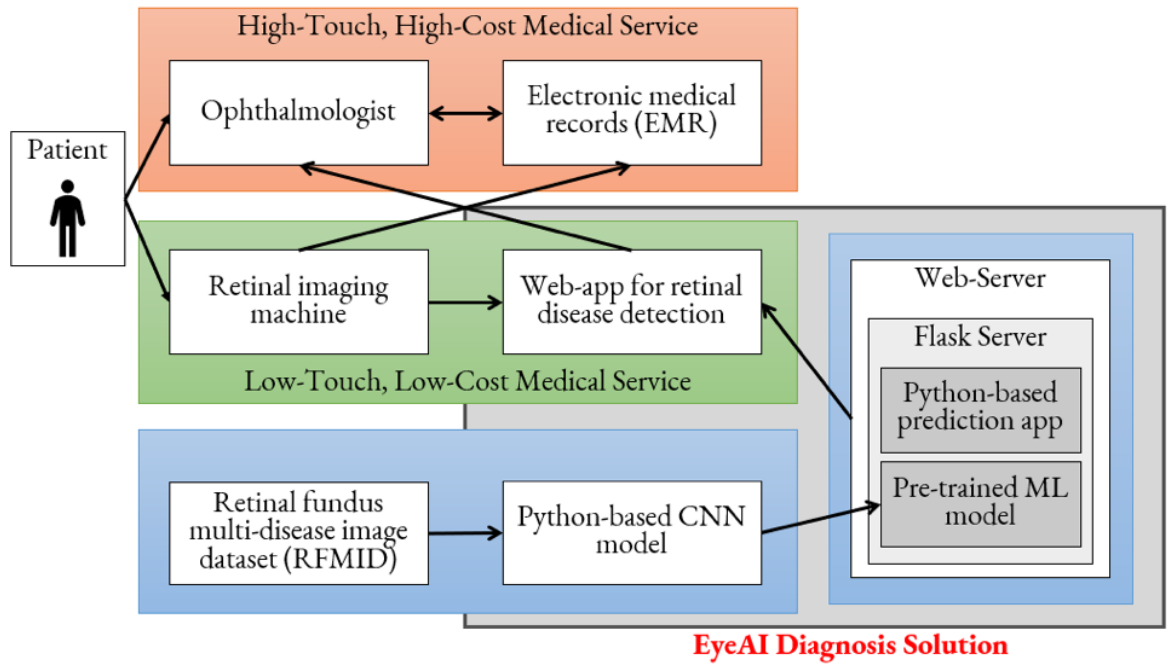

Figure 1. As an alternative to the traditional high-touch, high-cost medical service, patients can receive a diagnosis through a low-touch, low-cost service made possible by the EyeAI system. Patients obtain retinal fundus images using a retinal imaging machine, upload them to a web application, and receive a diagnosis made by the CNN model. The figure introduces EyeAI, which can transform ocular disease diagnosis by combining machine learning, web technologies, and existing imaging infrastructure to deliver efficient, low-cost, scalable care, reducing reliance on high-cost medical professionals.

Traditional ophthalmologic care involves an ophthalmologist's recommendation of a retinal imaging service to obtain retinal fundus (the back of the eye) images for a patient. After these images are processed and stored, the ophthalmologist will analyze the patient's retinal fundus images and diagnose the patient as having a diseased or a healthy retina. This information will then be stored in the electronic medical record (EMR). This scenario illustrates a high-touch, high-cost medical service that typically spans several weeks to several months. On the other hand, the proposed treatment option, EyeAI, enables patients to obtain retinal fundus images and receive a pre-diagnosis of the disease even before scheduling an appointment with an ophthalmologist. Using the web application for retinal disease detection, patients can upload an image and obtain a diagnosis indicating whether they have a diseased or healthy retina. This information can be transferred to an ophthalmologist. In the proposed system, the web application will run on a Flask-based web server that uses a pre-trained machine learning convolutional neural network (CNN) model to make predictions using Python. The machine learning model is trained on the Retinal Fundus Multi-Disease Image Dataset (RFMID).[8] This medical service, provided through EyeAI, can enable greater productivity, affordability, and accessibility of ocular disease diagnosis.

### *Novelty*

Prior AI-based ocular disease detection studies have primarily emphasized diagnostic accuracy while often overlooking practical aspects such as efficiency, accessibility, and broad disease coverage. Many models have focused on a limited set of ocular conditions and lacked deployment pathways suitable for real-world use. For example, Mostafa et al. developed a binary classification CNN that achieved approximately 98% accuracy but was limited to detecting disease within 7 ocular disease categories and was not publicly accessible for clinical or consumer use.[9] In contrast, EyeAI addresses these gaps by supporting binary classification across 45 ocular diseases and offering a low-cost, web-based interface designed for widespread accessibility. This focus on real-time, scalable deployment, particularly in underserved communities, distinguishes EyeAI from prior research and aligns with pressing needs for equitable diagnostic tools.[9-13] The efficiency and accessibility of diagnosis are increasingly crucial in the status quo, where productivity is vital, and equity is a priority.

Minimizing the need for frequent and expensive visits to ophthalmologists, EyeAI reduces overall healthcare expenses for patients and healthcare systems. Inexpensive retinal imaging machines and a pre-trained CNN model, when integrated into a web-based application, facilitate the financial feasibility of deploying the EyeAI system in low-income communities. EyeAI further bridges the gap in healthcare access by providing diagnostic tools to regions lacking sufficient medical infrastructure or physician care, helping overcome ophthalmologic care deserts. The early detection and treatment of ocular diseases enable the prevention of visual impairment and an improvement in quality of life.

## Methods

Following the frameworks put forth by prior research on AI-based ocular disease detection, a quantitative research study was performed involving the collection and processing of retinal fundus data, the construction of an adequate machine learning model to suit the research goals, and the training, testing, and refinement of the model.[9-13] Upon final model training and testing, the model was integrated into the web application for public use.

### *Data*

Data from the RFMID was used to develop a CNN model to classify retinal fundus images as diseased or healthy. It was created and compiled by medical professionals and researchers in 2021, and each image was approved for use by the Food and Drug Administration (FDA).[8] The RFMID consists of 3200 patient retinal fundus images (Figure 2) within 46 categories, 45 of which contain diseased retinal fundi.[8] The most prevalent diseases in the analyzed dataset were diabetic retinopathy, media haze, optic disk cupping, tessellation, drusen, and myopia (Figure 3).

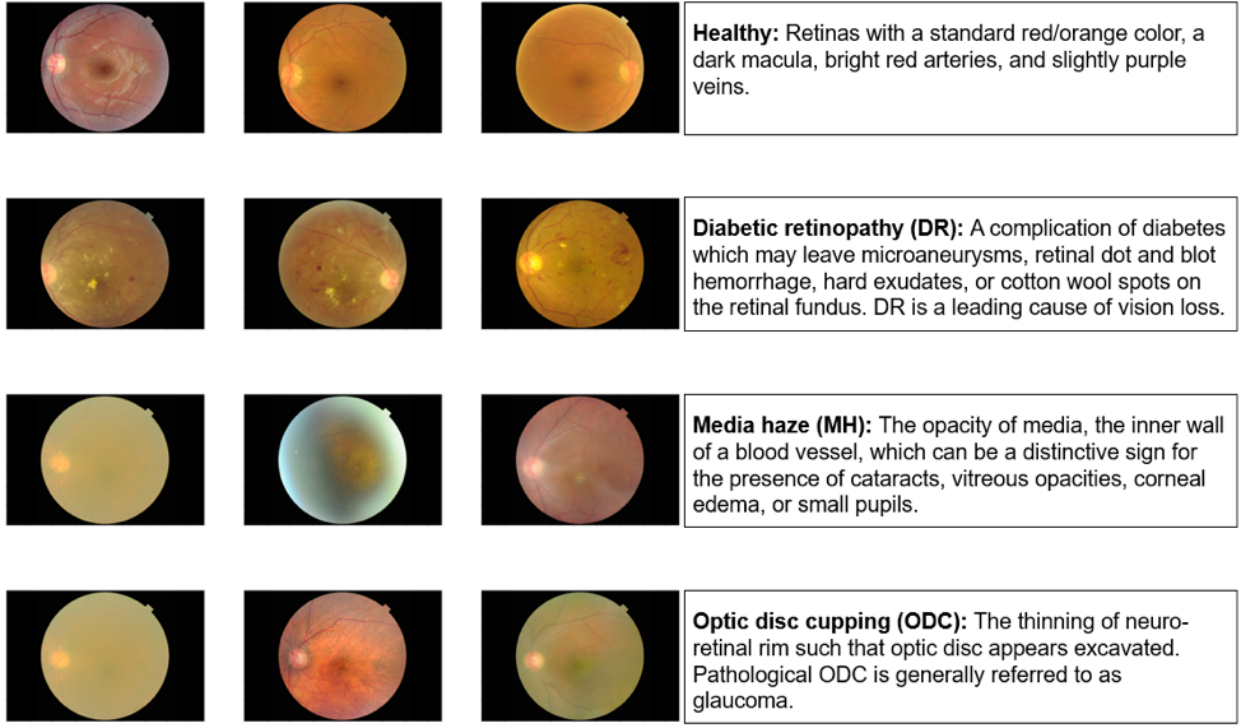

Figure 2. Examples of classified retinal fundus images from the RFMID, accompanied by disease descriptions, are shown. The four classifications depicted include eyes with a healthy retina and eyes with a retina affected by either diabetic retinopathy, media haze, or optic disc cupping, which are the three most common diseases in the RFMID. This helps visualize the degree to which retinal fundus images vary and how subtle differences can indicate the presence of a disease.

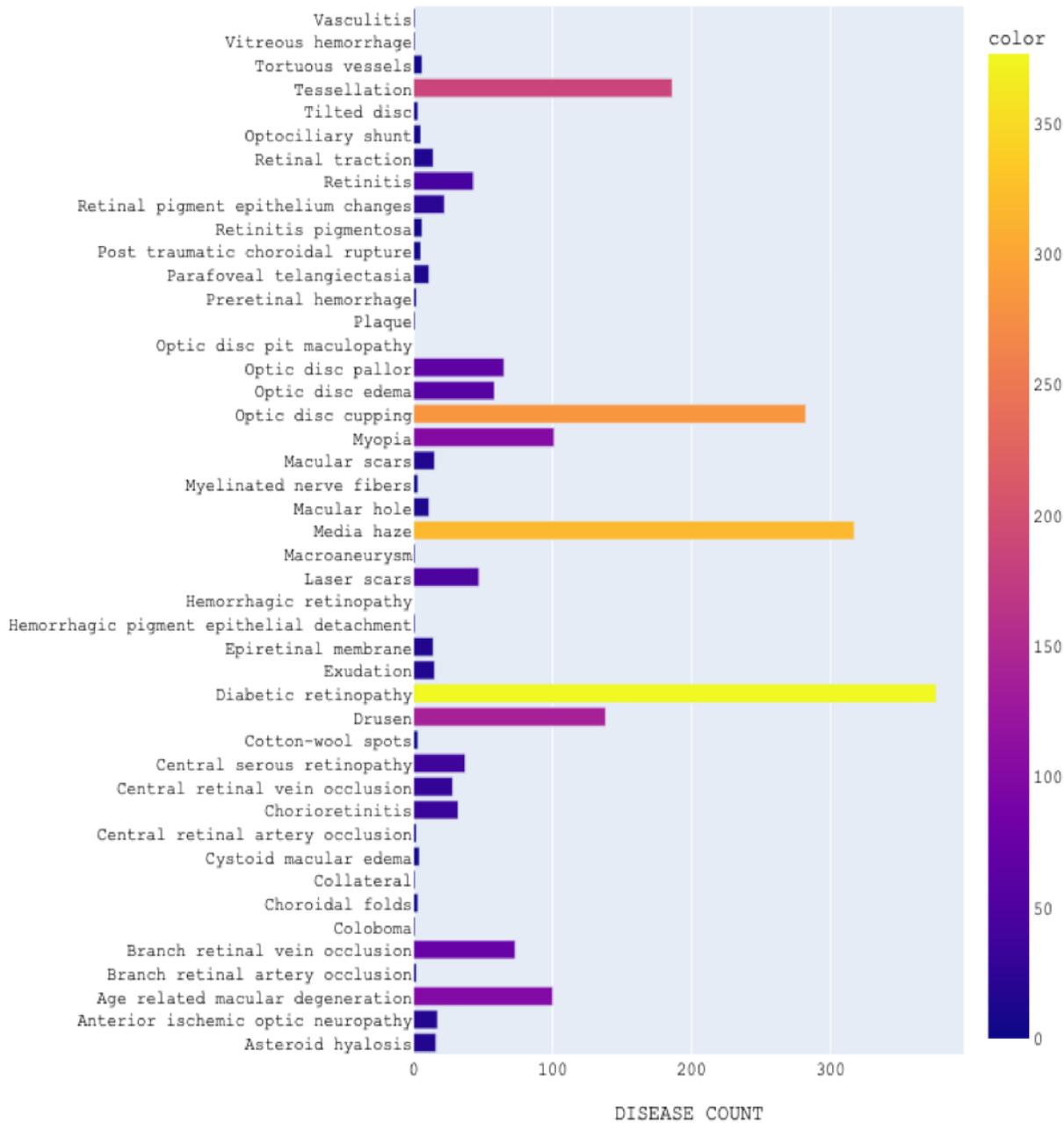

Figure 3. The distribution of various retinal conditions in the RFMID, with the number of cases (DISEASE_COUNT) on the x-axis, is depicted. Color intensity corresponds to disease frequency, as indicated by the color scale. The disease categories containing the most image data were diabetic retinopathy, media haze, optic disk cupping, tessellation, drusen, and myopia.

The RFMID data was split into training, validation, and testing datasets for use by the CNN. 60% of the data was used for training the model (n = 1920), 20% for validating the model (n = 640), and 20% for final testing of the completed model (n = 640). Given the size of the dataset, the 60/20/20 split offers a practical balance between computational efficiency, training amount, hyperparameter tuning, and evaluation rigor. The data was pre-processed using image augmentation through the TensorFlow

Keras library ImageDataGenerator object, which also helped in handling class imbalance, therefore improving model performance.

## *CNN Model*

A CNN machine learning model is used to analyze and classify image data based on features within the images; image classification takes an image input and determines an output, which, in this case, is the categorization as diseased or healthy. CNN models commonly include convolutional layers with the rectified linear unit (ReLU) activation function and pooling layers. These layers process the extensive data contained in pixels of an image to detect prominent features.

First, an image is processed into a matrix of pixels (using NumPy), with each pixel assigned a value for its red, green, and blue components (RGB), each on a scale of 0 to 255. RGB values define the pixel, and calculations are made based on these values. Each image has a resolution of 150 pixels by 150 pixels, meaning 22,500 pixels are processed for each image, and a total of 72,000,000 pixels are processed for all 3,200 images. The matrix for each image will undergo a series of transformations, or layers, to aggregate the information in the image.

The first layer that appears in a CNN is a convolutional layer. This layer applies a filter (a 3x3 matrix of values) over the receptive field (the pixel matrix). Each value of the filter is multiplied by the corresponding value of the receptive field; these values are added up, and the result is stored in a feature map. The filter is applied along the rows and columns of the receptive field, and the feature map containing aggregated data is stored.

Next, ReLU layers are activation functions applied to feature maps to account for images' non-linearity. With the activation function, all negative numbers in a feature map are set to 0. ReLU layers help normalize data by reducing the spread of information.

Additionally, the CNN model employs max-pooling layers to divide a feature map into non-overlapping areas of 2x2. Then the maximum value in each region is stored in a pooled feature map. This helps account for the variability of the images and further aggregates data.

Five sequences of a convolutional layer with ReLU and a max-pooling layer are added to the machine learning model to determine specific, prominent input characteristics and condense the information in the retinal fundus images (Figure 4). Finally, a binary output of "diseased" or "healthy" can be produced using a flattening layer. The complete CNN model architecture is depicted in Table 1, and the CNN training process is detailed in Table 2.

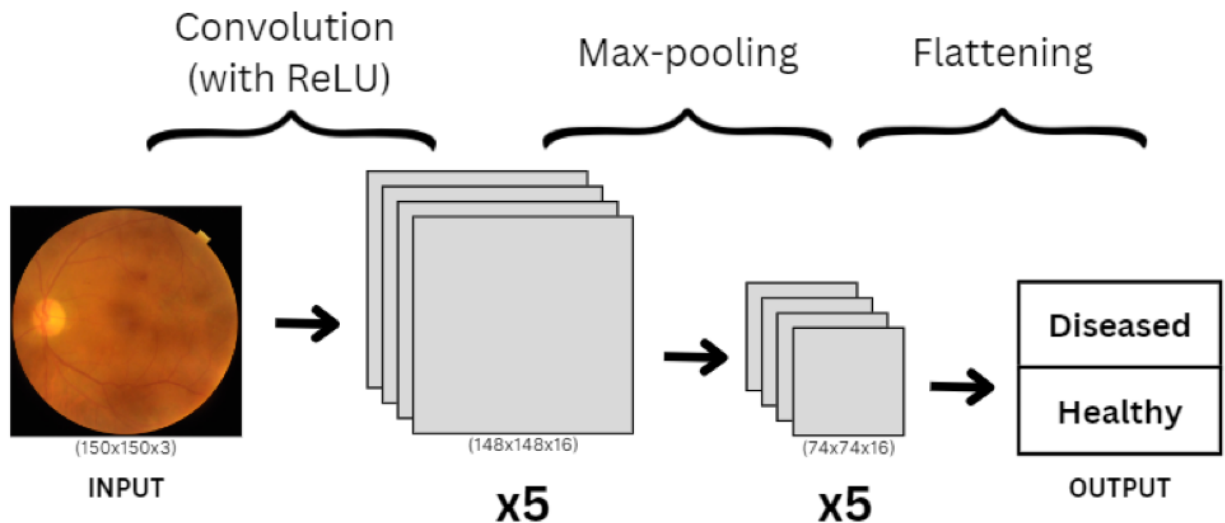

Figure 4. Data processing with the CNN model begins with a 150×150 color retinal fundus image input, which passes through five sequences of a convolutional layer with ReLU activation function, followed by a max-pooling layer. The feature maps are then flattened and fed into a fully connected layer to classify the image as "Diseased" or "Healthy." Through each of these layers, the data in the input image is processed and condensed to a final binary output.

Table 1. The CNN model architecture is detailed, including the algorithms used to create the model and their exact hyperparameters. A total of 14 layers are used to process the image from the input layer and produce a binary classification output layer.

| Layer | Type | Details |
| --- | --- | --- |
| Input Layer | Input | Input Shape: 150 × 150 × 3 |
| First Convolutional Layer | Conv2D | Filters: 16, Kernel Size: 3 × 3, Activation: ReLU |
| Pooling Layer 1 | MaxPooling2D | Pool Size: 2 × 2 |
| Second Convolutional Layer | Conv2D | Filters: 32, Kernel Size: 3 × 3, Activation: ReLU |
| Pooling Layer 2 | MaxPooling2D | Pool Size: 2 × 2 |
| Third Convolutional Layer | Conv2D | Filters: 64, Kernel Size: 3 × 3, Activation: ReLU |
| Pooling Layer 3 | MaxPooling2D | Pool Size: 2 × 2 |
| Fourth Convolutional Layer | Conv2D | Filters: 64, Kernel Size: 3 × 3, Activation: ReLU |
| Pooling Layer 4 | MaxPooling2D | Pool Size: 2 × 2 |
| Fifth Convolutional Layer | Conv2D | Filters: 64, Kernel Size: 3 × 3, Activation: ReLU |
| Pooling Layer 5 | MaxPooling2D | Pool Size: 2 × 2 |
| Flatten Layer | Flatten | N/A |
| First Fully Connected Layer | Dense | Units: 512, Activation: ReLU |
| Output Layer | Dense | Binary classification, Activation: Sigmoid |

Table 2. The specifications of the CNN model training process are detailed, providing an exact description of how the model was trained on the RFMID data. These training metrics were found to be effective in yielding a higher-performing CNN model.

| Attribute | Details |
| --- | --- |
| Loss Function | Binary Crossentropy |
| Optimizer | RMSprop |
| Learning Rate | 0.001 |
| Steps per Epoch | 45 |
| Epochs | 10 |
| Validation Steps | 8 |

### *Web Application*

This CNN model was saved to a hierarchical data format (HDF5) file. HTML and CSS files were created as a frontend user interface for uploading a retinal fundus image and receiving a diagnosis. The Flask framework was then used to link a central Python file to the model and the HTML and CSS pages. The image input to the web application is sent to the model, which outputs a diagnosis that is then sent back to the web application.

The web-based EyeAI application, accessible at https://eyeai-ci7emj7x5a-uc.a.run.app/, enables patients and healthcare providers to upload retinal fundus images and receive diagnostic results without requiring specialized software or extensive technical knowledge. Patients in remote or rural areas can benefit from advanced digital diagnostic tools without needing to travel to urban centers. The application facilitates rapid diagnoses, thereby reducing patient wait time. The online nature of the EyeAI system enables seamless integration with electronic medical records (EMR) of healthcare systems, allowing for efficient data transfer and storage.

## Results and Discussion

### *Model Performance*

After developing the CNN system, the EyeAI model was evaluated for statistical significance using validation data and tested using testing data. To provide an overall measure of the model's diagnostic performance, the accuracy of the model was investigated as a key metric. The area under the receiver operating characteristic curve (AUROC), which graphs the true positive rate (sensitivity) of the model against the false positive rate (1-specificity) of the model, was another metric utilized to evaluate the model due to its reflection of the model's discriminative ability. Precision, recall, and F1-score were also key metrics for evaluating the model, especially given the class imbalance.

The machine learning model achieved an accuracy of 80% and an AUROC of 0.698, demonstrating modest strength in image classification. The model achieved a high recall (sensitivity) of 91.82% but a low specificity of 7.78%, indicating a considerable ability to detect disease but with many false alarms on healthy inputs. Practically, this means that a predicted healthy diagnosis is a very strong indicator of a truly healthy input, but a predicted diseased input could correspond to either a healthy or diseased case. In a real-world medical environment, this could potentially lead to unnecessary referrals, contributing to patient stress and straining clinical resources.

This low specificity likely stems from the significant class imbalance within the RFMID dataset, the validation set of which contains only 90 healthy images compared to 550 diseased images spanning 45 distinct conditions. With such a wide range of disease types and relatively few healthy examples, the model may be overfit to the diverse diseased classes and misclassify healthy inputs, especially those that appear visually ambiguous. This tradeoff—high recall at the cost of specificity—is typical in imbalanced medical datasets, but it limits EyeAI's immediate clinical applicability.

To address this, future improvements should focus on increasing the diversity and volume of healthy training data, applying class weighting or focal loss during model training to penalize false positives more effectively, and exploring advanced approaches such as hierarchical classification to better distinguish subtle differences between healthy and diseased images. A relatively high precision of 85.88% and an F1-score of 0.8876 indicate moderate overall performance and reinforce the model's promise. However, substantial improvement in specificity is essential to reduce false alarms and enhance EyeAI's utility as a reliable diagnostic support tool. A confusion matrix for the model is shown in Table 3.

Table 3. A confusion matrix comparing the predicted outputs of the CNN model with the actual classifications of the retinal fundus images is shown. The confusion matrix highlights the number of true negatives, false negatives, true positives, and false positives. Notably, the number of true positive outcomes was vast, demonstrating the class imbalance in the RFMID.

|  | **Predicted Healthy** | **Predicted Diseased** |
| --- | --- | --- |
| **Actual Healthy (n = 90)** | true negative (TN): 7 | false positive (FP): 83 |
| **Actual Diseased (n = 550)** | false negative (FN): 45 | true positive (TP): 505 |

In the context of disease diagnosis in the health industry, false positives (FP) and false negatives (FN) have significant implications. False positives occur when the model incorrectly predicts ocular disease in a healthy individual. False positives may exacerbate unnecessary anxiety and potentially be costly for patients if they schedule follow-up tests; however, they do not provoke immediate health consequences that could devastate life. Still, minimizing false positives is essential to avoid adding undue complexity to the healthcare system. False negatives occur when the model fails to detect a disease that is present. False negatives can result in delayed diagnoses, delaying treatment, and possibly leading to the progression of the disease. Timely detection of ocular disease is crucial for preventing vision loss or severe eye complications. Thus, reducing the number of false negative outcomes is a priority.

### Model Selection

With numerous options for image classification models, a CNN was selected to develop the EyeAI system. Other machine learning models for image classification include transfer learning models, support vector machines (SVMs), random forest models, and Naive Bayes models. Transfer learning models utilize pre-trained neural networks on large datasets, requiring more computational resources and fine-tuning. Compared to SVMs, which are effective for smaller, well-defined datasets, CNNs scale better with larger image data and exhibit superior performance in feature extraction. While robust for various data types, random forest models often fall short in handling the complexity and volume of image data as efficiently as CNNs. Based on Bayes' theorem and assuming feature independence, Naive Bayes models are generally simpler and faster but do not capture the spatial hierarchies and complex patterns in image data as effectively as CNNs.

A CNN was chosen over the other models as it can automatically learn the features of images without requiring manual feature engineering. Additionally, CNNs require fewer parameters due to the standard filter used throughout each convolutional layer, enabling efficiency in model building. Furthermore, CNNs can easily manage large images, making their use for this study optimal. These factors, along with the generally high accuracy and speed associated with CNN models, led to the use of a CNN for EyeAI.

## Conclusion

The novel EyeAI System enables more accessible, scalable, and efficient ocular disease diagnosis. EyeAI should be implemented in areas where ophthalmologists are scarce, thereby allowing the diagnosis of ocular diseases for future treatment and care. Furthermore, EyeAI provides a secondary diagnostic opinion, enhancing diagnostic confidence and accuracy, and improving patient outcomes while increasing productivity within the healthcare system; EyeAI should be utilized throughout ophthalmology centers to aid in the diagnosis of ocular diseases. Improving accessibility to ocular disease diagnosis by integrating EyeAI in public spaces, such as wellness centers or clinics, can increase patient awareness of eye health and reduce visual impairments.

One of the primary limitations of the EyeAI study is the size and heterogeneity of the RFMID. The skewed classification of diseased retinal fundus images impacted the model's predictive performance. A more balanced dataset with an even representation of healthy and diseased images could improve the model's diagnostic accuracy. Furthermore, there was a clear skew in the distribution of images amongst disease categories, with certain diseases (e.g., diabetic retinopathy) being far more prevalent in the RFMID than others. This might impact the ability of the model to learn features from less common diseases, ultimately leading to misclassifications of retinal fundus images depicting those diseases. Addressing data collection limitations imposed by regulations such as the Health Insurance Portability and Accountability Act (HIPAA) and obtaining more retinal fundus images from FDA-approved sources will be crucial for advancing the model's capabilities. Further testing of various image classification models and parameters could also improve the EyeAI model.

Applying the EyeAI model to retinal imaging machines would result in even more efficient ocular disease diagnoses. Retinal fundi can be scanned using a retinal imaging machine, and a diagnosis can be returned almost instantaneously on the device. This would make the process of diagnosing ocular disease more seamless and streamlined.

By addressing the critical need for affordable and accessible ocular disease diagnosis, EyeAI can help revolutionize ocular healthcare and contribute to global health equity.

## Acknowledgements

Binita Patel is a STEM educator at FCS Innovation Academy, where she supports proficiency in IT. Ms. Patel's feedback has been essential throughout this project.

I would also like to thank Boehringer Ingelheim and 21st Century Leaders for their recognition and financial award, which supports this research.

## References

(1) World Health Organization. *World Report on Vision*. https://www.who.int/publications/i/item/9789241516570 (accessed 2025).

(2) Centers for Disease Control and Prevention. *Vision loss and diabetes*. https://www.cdc.gov/diabetes/diabetes-complications/diabetes-and-vision-loss.html (accessed 2024).

(3) International Diabetes Federation. *Global Diabetes Data & Statistics*. IDF Diabetes Atlas. https://diabetesatlas.org/ (accessed 2025).

(4) American Diabetes Association. *Eye health and diabetes*. https://diabetes.org/health-wellness/eye-health (accessed 2024).

(5) Centers for Disease Control and Prevention. *About vision impairment and falls among older adults*. https://www.cdc.gov/vision-health/prevention/older-adult-falls.html (accessed 2024).

(6) World Health Organization. *Ageing and health*. https://www.who.int/news-room/fact-sheets/detail/ageing-and-health (accessed 2025).

(7) Resnikoff, S.; Lansingh, V. C.; Washburn, L.; Felch, W.; Gauthier, T.-M.; Taylor, H. R.; Eckert, K.; Parke, D.; Wiedemann, P. Estimated number of ophthalmologists worldwide (International Council of Ophthalmology update): will we meet the needs? *British Journal of Ophthalmology* **2019**, *104* (4), 588–592. https://doi.org/10.1136/bjophthalmol-2019-314336.

(8) Pachade, S.; Porwal, P.; Thulkar, D.; Kokare, M.; Deshmukh, G.; Sahasrabuddhe, V.; Giancardo, L.; Quellec, G.; Mériaudeau, F. Retinal Fundus Multi-Disease Image Dataset (RFMID): a dataset for Multi-Disease detection research. *Data* **2021**, *6* (2), 14. https://doi.org/10.3390/data6020014.

(9) Mostafa, K.; Hany, M.; Ashraf, A.; Mahmoud, M. A. B. *Deep Learning-Based Classification of Ocular Diseases Using Convolutional Neural Networks*; Institute of Electrical and Electronics Engineers, 2023; pp 446–451. https://doi.org/10.1109/imsa58542.2023.10217787.

(10) Acevedo, E.; Orantes, D.; Acevedo, M.; Carreño, R. Identification of eye diseases through deep learning. *Diagnostics* **2025**, *15* (7), 916. https://doi.org/10.3390/diagnostics15070916.

(11) Vidivelli, S.; Padmakumari, P.; Parthiban, C.; DharunBalaji, A.; Manikandan, R.; Gandomi, A. H. Optimising deep learning models for ophthalmological disorder classification. *Scientific Reports* **2025**, *15* (1). https://doi.org/10.1038/s41598-024-75867-3.

(12) Marouf, A. A.; Mottalib, M. M.; Alhajj, R.; Rokne, J.; Jafarullah, O. An Efficient Approach to Predict Eye Diseases from Symptoms Using Machine Learning and Ranker-Based Feature Selection Methods. *Bioengineering* **2022**, *10* (1), 25. https://doi.org/10.3390/bioengineering10010025.

(13) Gulshan, V.; Peng, L.; Coram, M.; Stumpe, M. C.; Wu, D.; Narayanaswamy, A.; Venugopalan, S.; Widner, K.; Madams, T.; Cuadros, J.; Kim, R.; Raman, R.; Nelson, P. C.; Mega, J. L.; Webster, D. R. Development and validation of a deep learning algorithm for detection of diabetic retinopathy in retinal Fundus photographs. *JAMA* **2016**, *316* (22), 2402. https://doi.org/10.1001/jama.2016.17216.

## Authors

Shiv Garg studies cybersecurity and biotechnology and will major in computer science at the Georgia Institute of Technology starting in Fall 2025. He has a Google Data Analytics Professional Certificate and has received innovation awards from Broadcom, Boehringer Ingelheim, IEEE, CDC, and ACS. Shiv utilizes technology to serve his community.

Ginny Berkemeier, Ph. D., teaches biology at FCS Innovation Academy, and her mentorship and guidance have been instrumental to this research. She served as an advisor on the project, providing feedback to facilitate the completion of testing and analysis.